\documentclass[11pt]{article}
\usepackage{epsfig}

\def\la{\langle}\def\ra{\rangle}
\def\be{\begin{eqnarray}}\def\bea{\begin{eqnarray}}
\def\ba{\begin{eqnarray}}
\def\ee{\end{eqnarray}}\def\eea{\end{eqnarray}}
\def\ea{\end{eqnarray}}
\def\ben{\begin{eqnarray}}\def\bitem{\begin{itemize}}
\def\een{\end{eqnarray}}\def\eitem{\end{itemize}}
\def\np{Nucl. Phys.}
\def\bi{\bibitem}
\begin{document}
\begin{titlepage}
%\centerline{\large Preliminary}

\begin{center}
\ \\
{\Large \bf  The Pion Velocity in Dense Skyrmion Matter}
\\
\vspace{.30cm}
\renewcommand{\thefootnote}{\fnsymbol{footnote}}
Hee-Jung Lee$^{a}$,
Byung-Yoon Park$^{b}$, Mannque Rho$^{c}$
and Vicente Vento$^{a}$

\vskip 0.30cm

{(a) \it Departament de Fisica Te\`orica and Institut de
F\'{\i}sica
Corpuscular}\\
{\it Universitat de Val\`encia and Consejo Superior
de Investigaciones Cient\'{\i}ficas}\\
{\it E-46100 Burjassot (Val\`encia), Spain} \\ ({\small E-mail:
Heejung.Lee@uv.es,
Vicente.Vento@uv.es})

\vskip 1ex

{(b) \it CSSM, University of Adelaide, Adelaide 5005, Australia}\\
{\it and }\\
{\it Department of Physics,
Chungnam National University, Daejon 305-764, Korea}\\
({\small E-mail:
bypark@cnu.ac.kr})

\vskip 1ex

{(c) \it Service de Physique Th\'eorique, CEA Saclay}\\
{\it 91191 Gif-sur-Yvette, France}\\
{\it and} \\
{\it Department of Physics, Hanyang University, Seoul 133-791, Korea} \\
({\small E-mail: rho@spht.saclay.cea.fr}) \\

\end{center}
\vskip 0.3cm

\centerline{\bf Abstract} We have developed a field theory
formalism to calculate $in$-$medium$ properties of hadrons within
a unified approach that exploits a single Lagrangian to describe
simultaneously both matter background and meson fluctuations. In
this paper we discuss the consequences on physical observables of
a possible phase transition of hadronic matter taking place in the
chiral limit. We pay special attention to the pion velocity
$v_\pi$, which controls, through a dispersion relation, the pion
propagation in the hadronic medium. The $v_\pi$ is defined in
terms of parameters related to the matrix element in matter of the
axial-vector current, namely, the in-medium pion decay constants,
$f_t$ and $f_s$. Both of the pion decay constants change
dramatically with density and even vanish in the chiral limit when
chiral symmetry is restored, but the pion velocity does not go to
zero, decreasing at most 10\% over the whole density range
studied. A possible pseudogap structure is indicated. \vskip 0.5cm

\vskip 0.3cm \leftline{Pacs: 12.39-x, 13.60.Hb, 14.65-q, 14.70Dj}
\leftline{Keywords: pion velocity, pion decay constants,
dense matter}
%\vspace{0.5cm}

\end{titlepage}

\renewcommand{\thefootnote}{\arabic{footnote}}
\setcounter{footnote}{0}
\section{Introduction}
The initial motivation for building high energy heavy ion
colliders was the quest for the quark gluon plasma. However, the
data and the theoretical developments have shown that the phase
diagram of the hadronic matter is far richer and more interesting
than initially thought. At high temperature and/or density,
hadrons are expected to possess properties that are very different
from those at normal conditions. Chiral symmetry, which under
normal conditions is spontaneously broken, is believed to be
restored under extreme conditions. The value of the quark
condensate $\langle \bar{q} q \rangle$ of QCD is an order
parameter of this symmetry, and is expected to drop as the
temperature and/or density of hadronic matter is increased. Since
the pion is the ``litmus" indicator for spontaneously broken
chiral symmetry, the various patterns in which the symmetry is
realized in QCD, will be directly reflected in the properties of
the pions in medium.

The most fundamental quantities governing the dynamics of the pion
are its mass $m_\pi$ and its decay decay constant $f_\pi$. These
quantities have been the subject of previous studies in the
formalism adopted here~\cite{LPMRV03,LPRV03}. The property we are
interested in is encoded in the pion dispersion relation in
medium, which involves, besides the mass, the so-called pion
velocity in medium, $v_\pi$. This allows us to gain more insight
into the real time properties of the system under extreme
conditions and enable us to analyze how the phase transition from
normal matter to deconfined QCD phase takes place $bottom$-$up$
from the hadronic side. This issue has been addressed recently in
heat bath~\cite{PT96,SS02a,HKRS03}.

At nonzero temperature and/or density, we have to take into
account that the Lorentz symmetry is broken by the medium. In the
dispersion relation for the pion modes (in the chiral limit)
 \begin{equation}
p_0^2 = v_\pi^2 |\vec{p}|^2,
 \label{disp_rel}
\end{equation}
the velocity $v_\pi$ which is 1 in free-space must depart from 1.
This may be studied reliably at least at low temperatures and at
low densities via chiral perturbation theory~\cite{PT96}. The
in-medium pion velocity can be expressed in terms of the time
component of the pion decay constant, $f^t_\pi$ and the space
component, $f_\pi^s$,~\cite{fpits}
\begin{equation}
\begin{array}{l}
\langle 0 | A^0_a | \pi^b(p) \rangle_{\mbox{\scriptsize in-medium}}
= i f^t_\pi \delta^{ab} p^0, \\
\langle 0 | A^i_a | \pi^b(p) \rangle_{\mbox{\scriptsize
in-medium}} = i f^s_\pi \delta^{ab} p^i.
\end{array}
\label{decay_consts}
 \end{equation}
The conservation of the axial
vector current leads to the dispersion relation (\ref{disp_rel})
with the pion velocity given by
 \be v_\pi^2 = f_\pi^s /f_\pi^t.\label{vpi1}
  \ee

Other authors~\cite{SS02a} define two decay constants, $f_t$ and
$f_s$, different from those of eq.~(\ref{decay_consts})), through
the effective Lagrangian,
\begin{equation}
{\cal L}_{\mbox{\scriptsize eff}} = \frac{f_t^2}{4} \mbox{Tr} (\partial_0
U^{\dagger} \partial_0 U ) - \frac{f_s^2}{4} \mbox{Tr} (\partial_i U^{\dagger}
\partial_i U ) + \cdots, \label{Leff}
 \end{equation}
where $U$ is an SU(2)-valued chiral field whose phase describes
the in-medium pion. In terms of these constants, the pion velocity
is defined by
 \be
 v_\pi=f_s/f_t.\label{vpi2}
 \ee

What we are interested in here is the pion velocity in dense
medium as one approaches chiral restoration at the critical
density $\rho_c$. Before entering into this matter, we briefly
review the status of $v_\pi$ at high temperature which is more
relevant to heavy-ion collisions and the early Universe. There are
two basically different predictions depending upon what degrees of
freedom are taken into account, namely at $T=T_c$, $v_\pi=0$ in
one version which might be referred to as ``standard" and
$v_\pi\approx 1$ in another version which is non-standard. This
stark difference is the motivation for our investigation of dense
matter.

Consider the scenario described by the two-flavor linear sigma
model where the relevant degrees of freedom in heat bath are the
pions $\vec{\pi}$ and a scalar $\sigma$. In this model, as the
temperature approaches the critical temperature $T_c$, chiral
symmetry becomes restored, the pions and the scalar join in an
$O(4)$ multiplet, the phase transition belonging to the $O(4)$
universality class. This is the standard scenario largely accepted
by the community in the field. It has been shown~\cite{SS02a} that
this scenario predicts that at $T=T_c$, the pion velocity must
vanish. The argument is simple. Elevating the isovector axial
chemical potential $\mu_A$ to an $U(1)$ gauge field -- a powerful
trick, it follows from the low-energy effective action of QCD that
the isovector axial susceptibility (ASUS) $\chi_A$ is proportional
to $f_t^2$ that figures in eq.~(\ref{Leff}) where $f_t$ is the
time component of the fully renormalized pion decay constant in
heat bath. While one cannot reliably compute $f_t$ at low orders,
one can use the information that the ASUS must equal the isovector
vector susceptibility (VSUS) $\chi_V$ at the phase transition. Now
one knows that $\chi_V|_{T=T_c}\neq 0$ from lattice measurements
which implies that $f_t|_{T=T_c}\neq 0$  and that $f_s|_{T=T_c}=0$
from a general consideration. Thus we have at $T=T_c$
 \be
v_\pi=f_s/f_t=0. \label{sigmamodel}
 \ee
This is a simple and unambiguous prediction of the standard linear
sigma model.

The situation is markedly different when other light degrees of
freedom are relevant. It has been shown~\cite{HKRS03} that when
light vector mesons are in the picture as predicted by the hidden
local symmetry theory with the ``vector manifestation" \`a la
Harada and Yamawaki~\cite{HY:PR} (referred to as HLS/VM), the pion
velocity is close to 1 at all temperatures near $T_c$. The crucial
element that leads to this result is that the HLS theory with the
vector mesons $\rho$ and the pions coupled gauge invariantly and
matched to QCD at the matching scale $\Lambda_M$ has the vector
manifestation fixed point at which the parametric mass of the
vector meson $M_\rho$ and gauge coupling $g$ vanish. To this fixed
point flows the system as one approaches the criticality (in
temperature, in density or in the number of flavors). To one-loop
order in a generalized chiral perturbation theory in the presence
of light vectors -- expected to be a good approximation near the
critical point, the VM assures that $\chi_A=\chi_V$ at $T=T_c$,
$(f_\pi^t, f_\pi^s)\rightarrow (0, 0)$ (with $f_\pi$'s defined in
(\ref{decay_consts})) as $T\rightarrow T_c$. As a
consequence~\footnote{The pion velocity in this scenario is not
exactly equal to 1. There is a small deviation from 1 due to the
fact that the heat bath violates Lorentz invariance by a small
amount at the matching point $\Lambda_M$. This has been
computed~\cite{lv}.}
 \be
v_\pi\sim 1.
 \ee

We now turn to the main objective of this paper, which is to see
whether there is a similar dramatic dependence of the pion
velocity on the degrees of freedom also in dense matter. At
present, although it has been established that hadronic matter at
high density flows to the VM fixed point~\cite{HKR}, describing
dense matter in HLS/VM is found not to be easy because of coupled
renormalization group equations in the renormalization scale and
in density. Up to date, no conclusive result on physical variables
(e.g., $f_\pi$, susceptibilities etc) has been obtained. We need
therefore to resort to other formalisms.

At low density, the standard chiral perturbation theory without
vector degrees of freedom should be applicable. The in-medium pion
decay constants have been computed to first order in the baryon
number density $\rho$~\cite{MOW01}. The results are $f^t_\pi =
f_\pi (1-0.26 \rho/\rho_0)$ and $f^s_\pi = f_\pi (1-1.23
\rho/\rho_0)$, where $\rho_0$ is normal nuclear matter density.
Note that the space component is highly suppressed with respect to
the time one. Therefore the in-medium pion velocity scales as
 \be
v^2_\pi = f^s_\pi/f^t_\pi \sim 1- \rho/\rho_0.
 \ee
The rapid drop of $f^s_\pi$ -- and hence the pion velocity --
suggests that leading order-order chiral perturbation theory
cannot be trusted for density near that of normal nuclear matter.
This observation calls for a careful re-examination of power
counting.

QCD sum rules have also been applied to calculate the in-medium
pion decay constants~\cite{Kim02,KO03}. Unfortunately in the
pseudoscalar axial-vector correlation function used, the
dimension-5 condensate in the nucleon which cannot be calculated
reliably is found to play the most important role in splitting the
time and space components of the pion decay constant. This renders
quantitative estimates difficult. For instance, a positive value
of the condensate requires $f_\pi^s/f_\pi^t > 1$ and makes the
pion mass tachyonic ~\cite{Kim02}, at odds with causality. In
ref.~\cite{KO03}, this problem was circumvented by taking the
in-medium pion mass as an {\it input parameter} and obtained a
negative condensate. It was found that the contribution of the
intermediate $\Delta$ states is important in getting a sensible
result. Using the input values for the in-medium mass in the range
139 MeV $< m_\pi^* < 159$ MeV, the authors in \cite{KO03} obtained
$f_\pi^s/f_\pi = 0.37 \sim 0.78$ and $f_\pi^t/f_\pi = 0.63 \sim
0.79$ and as a consequence the in-medium pion velocity in the
range of
 \be
1/3 <v_\pi<1.
 \ee

In this paper, we study the in-medium pion velocity in the
systematic field theory scheme developed in our recent
works~\cite{LPMRV03,LPRV03}. There, the Skyrme picture is adopted
to describe in a unified way both the pions and dense baryonic
matter. A contact with HLS/VM theory of Harada and Yamawaki was
suggested there. A static soliton solution having an FCC crystal
structure~\cite{Kl85,CJJVJ89,KS89,JWFJRW}~\footnote{A slightly
different approach -- but similar in spirit -- to a crystal
structure of dense matter was discussed by Diakonov and
Mirlin~\cite{diakonov}. These authors studied the behavior of
nuclear matter with density by implementing the nucleon-nucleon
interaction by means of collective coordinate quantization of the
skyrmion-skyrmion interaction and arrived at the conclusion that
at twice or three time nuclear matter density a crystalline
structure could arise. This calculation rends an additional
support that a crystalline structure may indeed figure at high
density, as has been shown within our
approach~\cite{LPMRV03,LPRV03}.}

is found to provide a classical description of dense baryonic
matter. The pion is incorporated as fluctuating fields on top of
this dense skyrmion matter. The interactions between the pions and
matter modify the pion effective mass and decay constants. Since
the background skyrmion matter breaks Lorentz invariance, the
resulting space and time components of the pion decay constant
become different.

In the following section, we describe the Skyrme model approach
used for investigating the properties of the pions in dense
matter. A possible pseudogap structure is described in Section 3.
In Sec.~4, we study the in-medium pion velocity by taking into
account the background matter effect up to the second order in the
interactions. We make a brief conclusion in Sec.~5.

\section{Model Lagrangian}
The aim of our series of investigations has been to develop the
full dynamics of dense matter from a unique Lagrangian, which we
have taken for simplicity of Skyrme type. Here, we use a modified
Skyrme model Lagrangian \cite{EL85,NSVZ80,GJS8687,BR91}, which
incorporates in an effective way the scale anomaly of QCD in terms
of a scalar dilaton field~\footnote{The role of the dilaton field
in our model can be better understood as an interpolating field of
the ``soft gluon" degree of freedom that locks to chiral symmetry
as described in e.g. \cite{BGLR}. The ``hard or epoxy gluon"
degree of freedom can be considered as integrated out with its
effects embedded in the coefficients of the Lagrangian.}:
\begin{equation}
{\cal L} = \frac{f_\pi^2}{4}\left( \frac{\chi}{f_\chi} \right)^2
\mbox{Tr} (\partial_\mu U^\dagger \partial^\mu U)
 + \frac12 \partial_\mu \chi \partial^\mu \chi
- \frac{m_\chi^2 f_\chi^2}{4} \left(
({\chi}/{f_\chi})^4(\mbox{ln}(\chi/f_\chi)-\textstyle\frac14)
 + \frac14 \right)
\label{L0}
\end{equation}
where $U=\exp(i\vec{\tau}\cdot\vec{\pi}/f_\pi) \in SU(2)$ and
$\chi$ is the scalar field. The parameters in eq.~(\ref{L0}) are
related to physical properties, $f_\pi$ is the pion decay constant
in free space, $f_\chi$ the corresponding quantity for the scalar
field and $m_\chi$ its mass. For simplicity we will work with
massless pions.

The Lagrangian (\ref{L0}) in the $B=0$ sector describes only
mesons and yields the following pion and scalar field solutions in
the vacuum,
\begin{equation}
U=1, \hskip 3em \chi=f_\chi.
\end{equation}
The fluctuations about this vacuum can be described by
\begin{equation}
U_\pi=\exp(i\vec{\tau}\cdot\vec{\varphi}/f_\pi),
\hskip 1em \mbox{ and } \hskip 1em
\chi = f_\chi + \tilde{\chi}.
\end{equation}
Expanded up to second order in $\varphi$ and $\tilde{\chi}$, the
Lagrangian (\ref{L0}) takes the form
\begin{equation}
{\cal L} = \textstyle
 \frac12 \partial_\mu \varphi_a \partial^\mu \varphi_a
+ \frac12 \partial_\mu \tilde{\chi} \partial^\mu \tilde{\chi}
- \frac12 m_\chi^2  \tilde{\chi}^2 + \cdots.
\label{L1}\end{equation}

This Lagrangian can also describe, \`a la Skyrme~\cite{Sk62},
baryons, i.e., the skyrmions, as topological solitons if a proper
stabilizing term, e.g., the Skyrme term
\begin{equation}
{\cal L}_{\mbox{\scriptsize sk}} = \frac{1}{32e^2} \mbox{Tr}
\left( [ U^\dagger \partial_\mu U,
U^\dagger \partial_\nu U]^2 \right)
\end{equation}
is added.
The winding number associated with the homotopy of the static
solution $U_0(\vec{r})$ is taken as the baryon number,
\begin{equation}
B = \frac{1}{24\pi^2} \varepsilon^{ijk}
\int d^3 r
\mbox{Tr} (
 U_0^\dagger \partial^i U_0
 U_0^\dagger \partial^j U_0
 U_0^\dagger \partial^k U_0 ).
\end{equation}

%%%%%%%%%%%%%%%%%%%%%%%%%%%%%%%%%%%%%%%%%%%%%%%%%%%%%%%%%%%%%%%%%%%%%
The numerical results are sensitively dependent on the parameter
values chosen for the Lagrangian. Given that the Lagrangian used
is highly approximate, our results cannot be taken at their face
values. However we believe that the qualitative structure of the
results can be trusted. In this work, we will take the pion decay
constant to be the empirical value, say, $\sim$ 93 MeV. As for the
parameters associated with the dilaton, we have no direct
empirical information: neither the mass nor its structure is
experimentally identified. In the literature~\cite{FTS95,Song97},
they are adjusted so that the model fits finite nuclei as well as
nuclear matter. The dilaton decay constant comes out to be $\sim
240$ MeV, while the its mass ranges widely from 0.5 GeV to 1.5
GeV. We shall favor the lower masses, i.e. $\sim 700$ MeV, as
proposed by several investigations \cite{BGLR,hatta,vento}
although we shall report results for the wide range.

The parameter $e$ of the Skyrme term that figures crucially in
stabilizing the skyrmion also plays an important role in setting
the scale of dense matter. Unfortunately it is not obvious how to
implement density dependence in this parameter. One may naively
interpret the Skyrme term as arising when the $\rho$ vector meson
is considered to be much heavier than the scale involved and so is
integrated out. This cannot be correct however because other
mesons can contribute in the heavy-mass limit and can destabilize
the soliton. Furthermore, this term cannot be representing the
effect of the $\rho$ vector in dense medium since the $\rho$ meson
mass drops due to the vector manifestation~\cite{HY:PR}. On the
other hand, the Skyrme term figures at extreme short distances as
in the process of the proton decay \`a la Rubakov~\cite{rubakov}.
Therefore it seems reasonable to think that the single Skyrme term
subsumes a large number of massive degrees of freedom having the
appropriate quantum numbers lying above the scale involved in the
process. This implies that the constant $e$ -- representing
short-distance physics -- must be varying only slowly as a
function of density. This is somewhat analogous to what happens in
nuclear physics where the coefficients of zero-range counter terms
are ``universal" in that they are more or less independent of the
density. See \cite{parketal} where such a nuclear physics
phenomenon is encountered. For our purpose here we simply take it
as a constant. This assumption can be lifted in a model in which
the heavy degrees of freedom are explicitly treated for soliton
structure as in hidden gauge symmetry theory. In fact the explicit
account of vector mesons in the skyrmion structure is found to be
crucial for the bound state description of the pentaquark
$\Theta^+$ in the soliton model~\cite{pentaquark}. Although we do
not have a proof, we believe that the explicit presence of the
vector mesons would not significantly modify the qualitative
feature we are finding at and above nuclear matter density.

In this work, we pick the value of $e$ to fit, with the model,
either the baryon mass spectrum~\cite{ANW83} or the axial vector
coupling constant $g_A$~\cite{JM83}. The former fit leads to
$e\sim 5.45$ (but in this case, $f_\pi$ cannot have its empirical
value, and comes out 64 MeV), while the latter fit produces
$e=4.75$ (with $f_\pi$ at the empirical value).

Two skyrmions are most attractive when they are relatively rotated
by an angle $\pi$ about an axis perpendicular to the line joining
their centers. Thus the lowest energy configuration of skyrmions
for a given number density is in the crystal phase with all the
neighboring skyrmions in the most attractive orientations.
Moreover, the ground state of the scalar field $\chi$ is shifted
from the constant value $f_\chi$ to a static field configuration
$\chi_0(\vec{r})$ in order to reduce the energy of the whole
skyrmion-scalar system.
\section{Phase Transitions}
\subsection{Chiral restoration}
We review and update the numerical results on skyrmion matter
given in ref.~\cite{LPRV03}. The aim here is to give the
background for the calculation of the pion velocity discussed in
the following section with the error committed in the scale factor
in \cite{LPRV03} corrected.

In Fig.1 we show the energy per skyrmion $E/B$ of the skyrmion
crystal as a function of the baryon number density $\rho$. The
quantities in the chiral symmetry broken phase characterized by a
non-vanishing $\langle \chi \rangle$ are drawn by filled symbols,
while those in the chiral symmetry restored phase with $\langle
\chi \rangle=0$ are presented by unfilled symbols. In the
numerical calculation, the size and energy of the skyrmion have
been scaled in units of $(ef_\pi)^{-1}$ and $6\pi^2 f_\pi/e$,
respectively. When the decay constants are fixed to $f_\pi=93$ MeV
and $f_\chi=240$ MeV, the numerical results depend only on the
parameter $\mu_\chi = m_\chi/ef_\pi$. We show the results for
three different values of $\mu_\chi$, namely 1.25 (by triangles),
1.63 (by circles) and 2.26 (by squares), corresponding to
$m_\chi=550$, 720 and 1000 MeV for $f_\pi=93$MeV and $e=4.75$. The
density is given in units of $0.015(ef_\pi)^3$, where the constant
0.015 is chosen so that it equals the normal nuclear density
$0.17$(fm)$^{-3}$ for $f_\pi=93$ MeV and $e=4.75$. We should
stress that the exact value of the density does not have a precise
physical meaning, since the density scales with $e$, the Skyrme
parameter, which as mentioned above, is not well defined. However
we believe it to be a useful, albeit qualitative, guide to the
magnitudes of the densities involved.

As the density increases we reach a value at which the chiral
symmetry restored phase has a lower energy than the chiral
symmetry broken phase, i.e., a chiral phase transition takes
place. Note that, even beyond the phase transition density,
$\langle \chi\rangle\neq 0$  is a {\em local minimum}. On the
other hand, in the presence of the background matter, $\langle
\chi \rangle =0$ is always a local minimum whereas in free space,
it was just an unstable extremum of $V(\chi)$. This phenomenon can
be easily understood using the approximation introduced in
ref.~\cite{LPRV03}, namely, taking $\chi(\vec{r})$ as a constant
$\chi_0$. In that approximation, the energy per baryon $E/B$
becomes
\begin{equation}
\begin{array}{rcl}
E/B &=&
(E_2/B) (\chi_0/f_\chi)^2 + E_4/B \\
&& + V(\chi_0) \times \mbox{[Volume occupied by a single skyrmion]},
\end{array}
\end{equation}
where $E_{2,4}/B$ are the contributions to the energy per baryon
of the quadratic and quartic in the Lagrangian (\ref{L0}) and
$V(\chi)$ is the potential energy density of the dilaton field.
Now, it is evident that
\begin{equation}
\left.
\frac{ \partial^2 (E/B)}{\partial (\chi_0/f_\chi)^2}
\right|_{\chi_0 = 0} = 2 (E_2/B) ,
\end{equation}
and the system is quasi-stable at $\chi_0 = 0$ as long as $2
(E_2/B) > 0$.  These local minima become true minima depending on
the density, the former phase for low densities and the latter for
densities above the phase transition.

%------------------------------ fig 1 --------------------------------------
\begin{figure}
\centerline{\epsfig{file=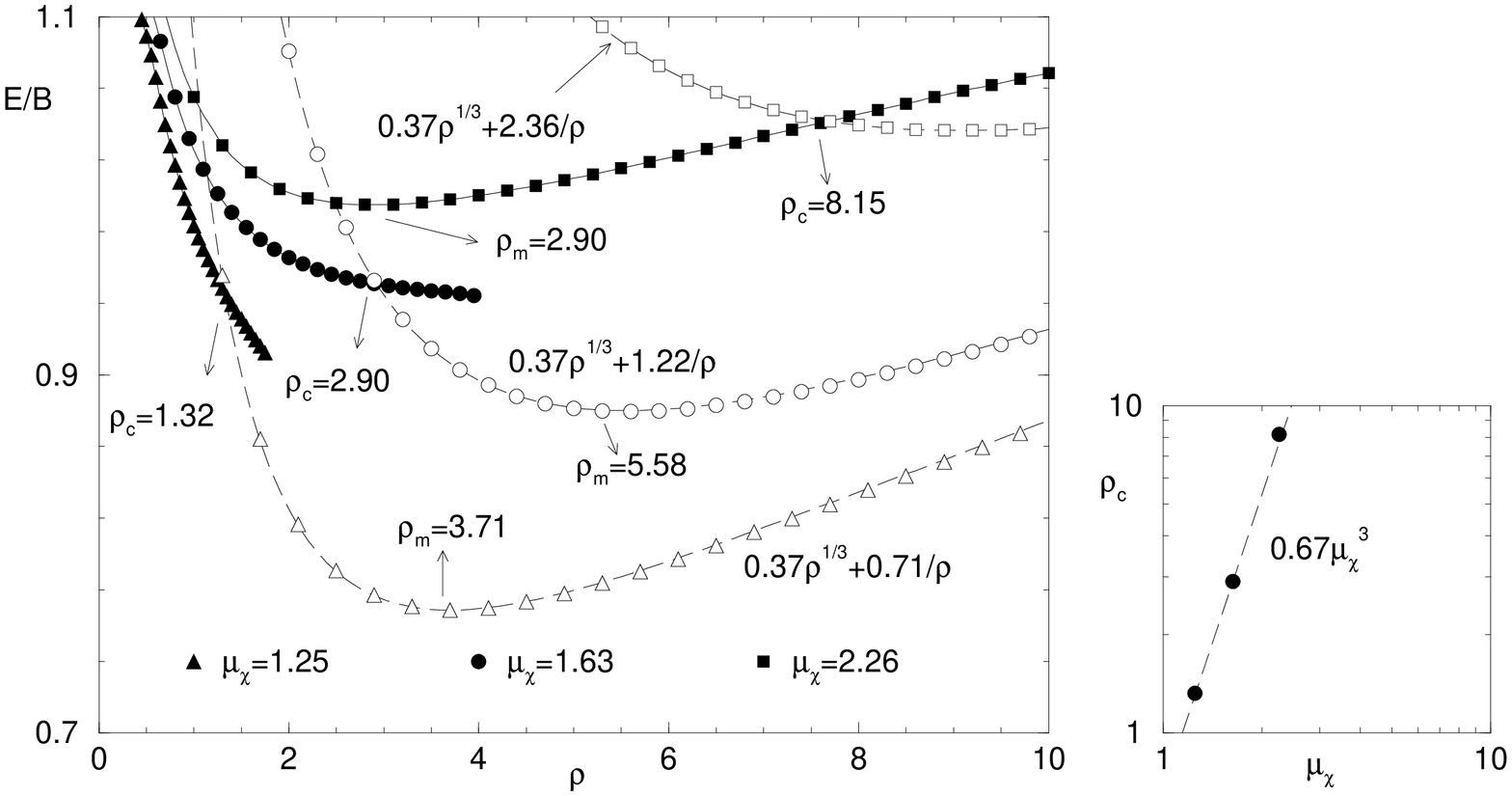,width=8cm,height=15cm,angle=270}}
\ \caption{$E/B$ as a function of density. Numerical data for the
$\langle \chi \rangle \neq 0$ phase are presented by filled
symbols and those for the $\langle \chi \rangle = 0$ phase are
presented by empty symbols. The density is in unit of
$0.015(ef_\pi)^3$, which equals normal nuclear matter density
$\rho_0=0.17$ (fm)$^{-3}$ when the conventional values, $e=4.75$
and $f_\pi=93$ MeV, are used. $E/B$ is given in unit of $6\pi^2
f_\pi/e$. \label{eperb}}
\end{figure}
%----------------------------- end fig 1 -----------------------------------

As an aside, we note that the numerical data for the $\langle
\chi\rangle = 0$ phase can be fit by a simple curve
\begin{equation}
E/B = a \rho^{1/3} + b / \rho,
\end{equation}
where the first term comes from the Skyrme term contribution $E_4/B$
that scales as $\rho^{1/3}$ and the second comes from the potential energy
of the dilaton field. The coefficient $a$ is almost
independent of $\mu_\chi$.

The chiral phase transition takes place at the density, $\rho_c$,
where the phase characterized by $\langle \chi \rangle=0$ and that
by $\langle \chi \rangle \neq 0$ have the same $E/B$. The critical
density depends sensitively on the parameter $\mu_\chi$. In Fig.
1, we show results for, $\mu_\chi=1.25$, 1.63 and 2.26 for which
$\rho_c $ becomes 1.32, 2.90 and 8.15, respectively. The figure
drawn in a small box is the critical density $\rho_c$ as a
function of $\mu_\chi$. Note that it scales as $\mu_\chi^3$.
The dashed line in the figure is not a fit but  $0.67\mu_\chi^3$
line just passing through the data point at $\mu_\chi=1.63$.

\subsection{Pseudogap phase}

\begin{figure}
\centerline{\epsfig{file=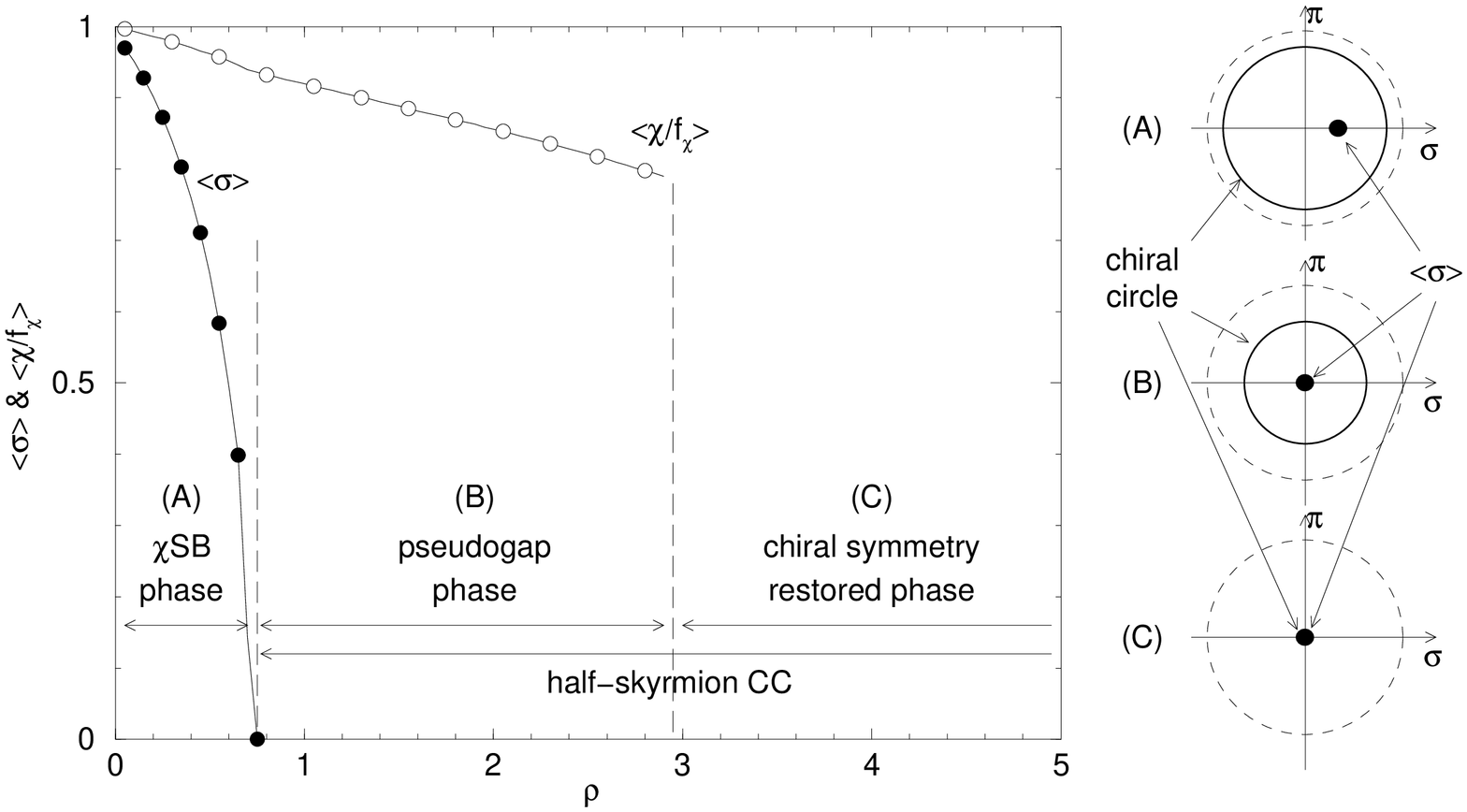,width=8cm,height=13.5cm,angle=270}}
\caption{Average values of $\sigma=\frac12\mbox{Tr}(U)$ and $\chi/f_\chi$
of the lowest energy crystal configuration at a given baryon number
density.}
\end{figure}

In Fig. 2 we show the average values of $\sigma =
\frac12\mbox{Tr}{U}$ and $\chi/f_\chi$ over space, which provide
direct information on the structure of the crystal and can be used
as order parameters of the phase transition. These data show that
there is a ``phase transition" before the chiral restoration
transition, at the density at which the expectation value of
$\sigma$ vanishes. Recalling that the expectation of $\sigma$ is
proportional to the quark condensate, this may indicate a change
in the chiral symmetry structure due to destructive interference
of the phase of the quark condensate with a nonvanishing modulus.
In the case of the massive pions, such a phase transition is not
abrupt but continuous and this is the reason why we failed to
notice it in our previous work~\cite{LPRV03}. We identify this
phase transition with $\la\sigma\ra=0$ while $\la\chi\ra\neq 0$ as
of a {\em pseudogap type}. This phase persists in an intermediate
density region, where the $\langle \chi/f_\chi\rangle$ does not
vanish while $\langle \sigma \rangle$ does~\cite{RD38W92}.
We denote the density
at which the pseudogap phase transition appears as $\rho_p$. Since
the vanishing of $\sigma$ is possible because each skyrmion maps
the chiral circle onto space in order to carry nontrivial baryon
number, it may be that the pseudogap phase is an artifact of the
skyrmion matter. We point out however that a similar pseudogap
structure was proposed in hot QCD~\cite{zarembo}.

The two step phase transition is schematically illustrated in
Fig.~2.
\begin{enumerate}
\item [(A)]
At low density ($\rho < \rho_p$), matter slightly reduces the
vacuum value of the dilaton field from that of the baryon free
vacuum. This implies a shrinking of the radius of the chiral
circle by the same ratio. Since the skyrmion takes all the values
on the chiral circle, the expectation value of $\sigma$ is not
located on the circle but inside the circle. Skyrmion matter at
this density is in the chiral symmetry broken phase.

\item [(B)]
At some intermediate densities ($\rho_p < \rho < \rho_c$), the
expectation value of $\sigma$ vanishes while that of the dilaton
field is still nonzero. The skyrmion crystal is in a CC
configuration made of half skyrmions localized at the points where
$\sigma=\pm 1$. Since the average value of the dilaton field does
not vanish, the radius of the chiral circle is still finite. Here,
$\langle \sigma \rangle =0$ does not mean that chiral symmetry is
completely restored. We interpret this as a pseudogap phase.

\item [(C)]
At higher density ($\rho > \rho_c$), the phase characterized by
$\langle \chi/f_\chi \rangle=0$ becomes energetically favorable.
Then, the chiral circle, describing the fluctuating pion dynamics,
shrinks to a point.

\end{enumerate}

In the case of massive pions, the chiral circle is tilted by the
explicit symmetry breaking term. Thus, the exact half-skyrmion CC,
that requires a symmetry between points with values $\sigma=+1$
and $\sigma=-1$ cannot be constructed and consequently the phase
characterized by $\langle \sigma \rangle=0$  cannot exist at any
density. However, $\langle \sigma \rangle$ is always inside the
chiral circle and its value drops much faster than that of
$\langle \chi/f_\chi\rangle$. Thus, if the pion mass is small
enough, a pseudogap phase can appear in the model.

\begin{figure}
\centerline{\epsfig{file=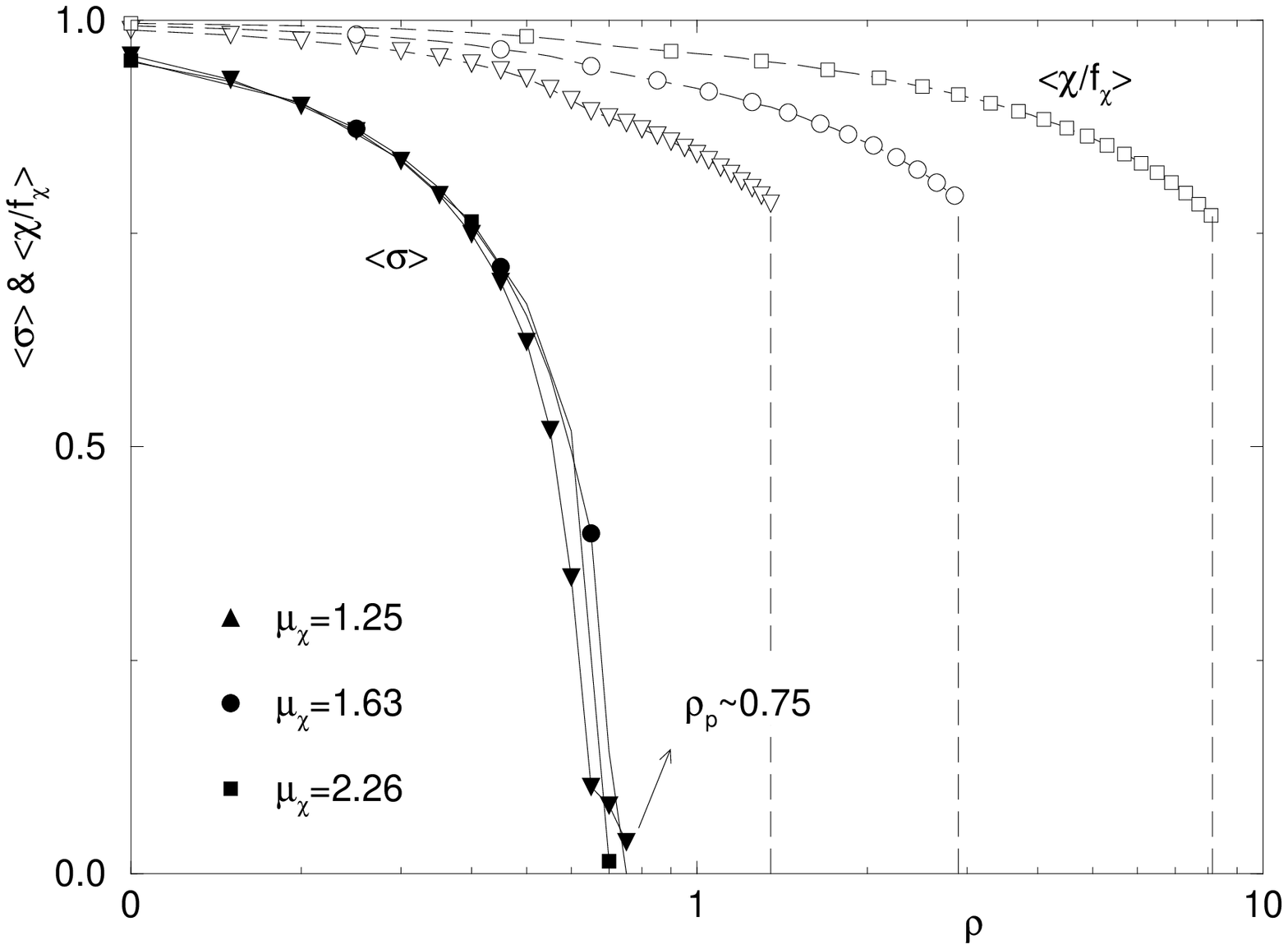,width=8cm,height=11cm,angle=270}} \
\caption{$\mu_\chi$ dependence of two step phase transitions.}
\end{figure}

In Fig. 3 we show the two step phase transition at various values
of the parameter $\mu_\chi$. Contrary to $\rho_c$, $\rho_p$
depends only insensitively on $\mu_\chi$. For small $\mu_\chi \sim
1$, $\rho_p \sim \rho_c$ and its value is  smaller than that for
large $\mu_\chi$. As $\mu_\chi$ increases, $\rho_c$ increases
proportionally to $\mu_\chi^3$ as was  seen in Fig. 1, while
$\rho_p$ saturates to $0.78$. Since the density unit is
$(ef_\pi)^3$ and recalling that $\sqrt2 ef_\pi \sim m_\rho$, we
can summarize the phenomena in terms of the parameters $e$ and
$m_\chi$ as,
\begin{enumerate}
\item At density $\rho_p \sim 0.78 \times 0.015(ef_\pi)^3 \sim 0.004 m_\rho^3$ ,
a pseudogap phase transition occurs.

\item
At density $\rho_c\sim 0.67\mu_\chi^3 \times 0.015(ef_\pi)^3 \sim
0.01 m_\chi^3$, the chiral phase transition takes place,
\end{enumerate}
with $m_\chi \sim m_\rho$.

%%%%%%%%%%%%%%%%%%%%%%%%%%%%%%%%%%%%%%%%%%%%%%%%%%%%%%%%%%%%%%%%%
\section{In-Medium Pion Velocity}

The pion and scalar in dense medium can be described by
fluctuations on skyrmion matter through the Ans\"{a}tze
\begin{equation}
U = \sqrt{U_\pi} U_0(\vec{r}) \sqrt{U_\pi},
\hskip 1em \mbox{ and } \hskip 1em
\chi = \chi_0(\vec{r})+\tilde{\chi}.
\end{equation}
Expanding up to the second order in the fluctuating
fields, we obtain
\begin{eqnarray}
{\cal L} &=& \textstyle
\frac12 G^{ab}(\vec{r})
  \partial_\mu \varphi_a \partial^\mu \varphi_b
+ \epsilon_{abc}  \varphi_a \partial_i \varphi_b V^{i}_c(\vec{r})
\nonumber\\
&& \textstyle +\frac12 \partial_\mu \tilde{\chi} \partial^\mu
\tilde{\chi} - \frac12 M(\vec{r}) \tilde{\chi}^2 + P^i_a(\vec{r})
\tilde{\chi} \partial_i \varphi_a. \label{L2}\end{eqnarray} This
Lagrangian describes {\em single particles} moving under ``local
background potentials" provided by the static field configurations
$U_0(\vec{r})$ and $\chi_0(\vec{r})$. We have shortened the
lengthy expressions for the local potentials introducing
$G^{ab}(\vec{r})$, $V^{i}_c(\vec{r})$, $ M(\vec{r})$,
$P^i_a(\vec{r})$. Their explicit forms are shown in
ref.~\cite{LPRV03}. Assuming background matter to be known,
eq.~(\ref{L2}) shows, when compared with eq.~(\ref{L1}),  that a
direct $\varphi$-$\chi$ (not only $\varphi^2$-$\chi$) coupling
might appear, which is absent in free space. Hereafter, we will
drop the tildes for the fluctuating scalar field.

In refs.~\cite{LPMRV03,LPRV03}, by replacing the local potentials
with their averages over space, we obtained
\begin{equation}
{\cal L}_{\mbox{\scriptsize eff}} = \textstyle \frac12 \langle
G^{aa} \rangle \partial_\mu \varphi_a \partial^\mu \varphi_a +
\frac12 \partial_\mu \chi \partial^\mu \chi - \frac12 \langle M
\rangle \chi^2 + \cdots. \label{Leff1}\end{equation} We interpret
this result as an effective Lagrangian for the in-medium pions and
scalar with ``intrinsic" parameters $\langle G^{aa} \rangle$ and
$\langle M \rangle$, which correspond to the in-medium pion decay
constant and the effective mass for the scalar respectively. In
this naive approximation, Lorentz symmetry is still preserved.
Thus, the pion velocity is 1 at this level. However, static local
potentials will definitely break the Lorentz symmetry. In this
work, we are going to evaluate the deviation of the pion velocity
from 1 due to this background.

Our strategy in performing the calculation is based on the
following result. When local interactions with background matter
are taken into account they lead to an effective Lagrangian for
pion dynamics in the form of eq.~(\ref{Leff}) which defines the
quantities necessary for the calculation. Let us be more specific.
As far as the single pion is concerned, to lowest order in the
fields,  eq.~(\ref{Leff}) with
$U=\exp(i\vec{\tau}\cdot\vec{\varphi}/f_\pi)$ becomes
%-------------------------- footnote --------------------------
\footnote{If one substituted
$U=\exp(i\vec{\tau}\cdot\vec{\varphi}^*/f_t)$
with {\em renormalized pion fields} $\varphi^*_a$,
one would have
$${\cal L}_{\mbox{\scriptsize eff}}
=  \frac12 (\partial_0 {\varphi}^*_a \partial_0 {\varphi}^*_a
- v_\pi^2
\partial_i {\varphi}^*_a \partial_i {\varphi}_a^* )
+\cdots .$$ }
%---------------------- end of footnote ------------------------
\begin{equation}
{\cal L}_{\mbox{\scriptsize eff}}
=  \frac12 \left(\frac{f_t}{f_\pi} \right)^2
\partial_0 {\varphi}_a \partial_0 {\varphi}_a
- \frac12 \left(\frac{f_s}{f_\pi} \right)^2
\partial_i {\varphi}_a \partial_i {\varphi}_a
+\cdots . \label{Leff0}\end{equation} Thus the  procedure consists
of calculating the propagators for the pion and the scalar fields
from  (\ref{L2}) by solving for the background potentials. Then,
matching these propagators with those obtained from the effective
Lagrangian (\ref{Leff0}), we have for the pion
\begin{equation}
\frac{1}{Z_t^{-1} p_0^2 - Z_s^{-1} \vec{p}^2}
\label{prop_eff}\end{equation}
where $Z_{t,s}^{-1}=(f_{t,s}/f_\pi)^2$.

Before proceeding to the calculation, we first obtain relations
between the parameters, $f_{t,s}$ and $f_\pi^{t,s}$, defined in
Introduction. The effective Lagrangian (\ref{Leff}) yields the
following axial vector current components
\begin{equation}
\begin{array}{l}
\displaystyle
A_0^a =-i\frac{f_t^2}{4}\mbox{Tr} \left(
  \partial_0 U U^\dagger \frac{\tau^a}{2}
- \partial_0 U^\dagger U \frac{\tau^a}{2} \right)
= ( {f_t^2}/{f_\pi} ) \partial_0 \varphi, \\
\displaystyle
A_i^a = -i\frac{f_t^2}{4}\mbox{Tr} \left(
  \partial_i U U^\dagger \frac{\tau^a}{2}
- \partial_i U^\dagger U \frac{\tau^a}{2} \right)
= ( {f_s^2}/{f_\pi} ) \partial_i \varphi,
\end{array}
\end{equation}
where the last equality is obtained by substituting
$U=\exp(i\vec{\tau}\cdot\vec{\varphi}/f_\pi)$. When the
expectation value of these axial vector current components is
taken as in eq.~(\ref{decay_consts}), we obtain the relations,
\begin{equation}
f_t = f_\pi^t, \mbox{ and } f_s = \sqrt{f_\pi^s f_\pi^t}.
\label{relation}\end{equation} Similar relations are derived in
ref.~\cite{HKRS03} for the real part of the complex pion decay
constants. In what follows, the results will be given in terms of
$f_t$ and $f_s$.

In order to obtain nontrivial results for the pion velocity in the
medium we need  to take into account the higher order effects of
the background potentials. We shall proceed here {\em in a
perturbative} scheme. To do so we  decompose the Lagrangian into
an unperturbed part, ${\cal L}_0$, and an interaction part, ${\cal
L}_I$. The simplest way to do so is to take the free Lagrangian
(\ref{L1}) as ${\cal L}_0$ and the rest as ${\cal L}_I$.
Explicitly,
%---------- L_I ---------------
\begin{eqnarray}
{\cal L}_{I} &=& \textstyle
\frac12 (G^{ab}(\vec{r})-\delta^{ab})
  \partial_\mu \varphi_a \partial^\mu \varphi_b
+ \epsilon^{abc}  \varphi_a \partial_i \varphi_b V^{i}_c(\vec{r})
\nonumber\\
&& \textstyle
- \frac12 (M(\vec{r})-m_\chi^2) \tilde{\chi}^2
+ P^i_a(\vec{r}) \tilde{\chi} \partial_i \varphi_a.
\label{Lint}
\end{eqnarray}

The presence of the local interaction potential
$G^{ab}(\vec{r})$ in the pion kinetic energy term
makes the quantization process of the fluctuating pions
nontrivial.
The conjugate momenta of the pion fields
$\varphi_a$ are given by
\begin{equation}
\Pi_a \equiv \frac{\partial {\cal L}_0}{\partial \dot{\varphi}_a}
 = \dot{\varphi}_a.
\end{equation}
We obtain
\begin{equation}
{\cal H}={\cal H}_0 + {\cal H}_I,
\label{H}\end{equation}
where
$$
{\cal H}_0 = \textstyle \frac12 ( \dot{\varphi}^a
\dot{\varphi}^a + \partial_i \varphi^a \partial_i \varphi^a )
+ \frac12(\dot{\chi}\dot{\chi} + \partial_i \chi\partial_i \chi
+ m_\chi^2 \chi^2 ),
\eqno(\mbox{\ref{H}a})$$
and
$${\cal H}_I = -{\cal L}_{I}.
\eqno(\mbox{\ref{H}b})$$
Note that the interaction Hamiltonian ${\cal H}_{I}$ is
simply $-{\cal L}_I$.
The free propagators defined by ${\cal H}_0$
and the interaction potentials appearing due to ${\cal H}_I$
are summarized in Fig.~4.
In Fig.~4, $G^{ab}(\vec{\ell})$, for example, is the Fourier
transformed of the local potential $G^{ab}(\vec{r})$:
\begin{equation}
G^{ab}(\vec{\ell}) = \frac{1}{V_{\mbox{\scriptsize box}} }
\int_{\mbox{\scriptsize box}} d^3 r e^{i\vec{\ell}\cdot\vec{r}}
G^{ab}(\vec{r}),
\end{equation}
where the integration is over a unit box of the crystal and
$V_{\mbox{\scriptsize box}}$ is its volume.
Due to the periodic structure of the crystal only discrete
momentum values are possible.

%------------------------  Figure 4 -------------------------------
\begin{figure}
\begin{center}
\setlength{\unitlength}{1mm}
\begin{picture}(150,65)
\linethickness{0.8pt}
%--------------------- propagators -----------------------
\put(29,62){\makebox(0,0)[b]{$\varphi_a$}}
\put(28,60){\circle*{1}}
\multiput(28,60)(5,0){7}{\line(1,0){3}}
\put(62,60){\circle*{1}}
\put(61,62){\makebox(0,0)[b]{$\varphi_b$}}
\put(45,57){\makebox(0,0)[c]{$(p_0,\vec{p})$}}
\put(45,48){\makebox(0,0)[c]{$\displaystyle\frac{\delta_{ab}}
{p_0^2-\vec{p}^2}$}}
\put(88,62){\makebox(0,0)[b]{$\chi$}}
\put(88,60){\circle*{1}}
\put(88,60){\line(1,0){34}}
\put(122,60){\circle*{1}}
\put(123,62){\makebox(0,0)[b]{$\chi$}}
\put(105,57){\makebox(0,0)[c]{$(p_0,\vec{p})$}}
\put(105,48){\makebox(0,0)[c]{$\displaystyle\frac{1}{p_0^2-\vec{p}^2
- m_\chi^2}$}}
%--------------------- pi-pi -----------------------
\put(30,25){\makebox(0,0)[c]{${\cal H}^{\varphi\varphi}_I(\vec{\ell})$}}
\put(20,26){\makebox(0,0)[b]{$\varphi_a$}}
\put(23,24){\makebox(0,0)[tr]{$(p_0,\vec{p})$}}
\put(40,26){\makebox(0,0)[b]{$\varphi_b$}}
\put(38,24){\makebox(0,0)[tl]{$(q_0,\vec{q})$}}
\put(23,25){\circle*{1}}
\put(30,25){\circle{14}}
\put(37,25){\circle*{1}}
\put(30,13){\makebox(0,0)[c]{$-(p_0^2-\vec{p}\cdot\vec{q})
(G^{ab}(\vec{\ell})-\delta^{ab})$}}
\put(30,7){\makebox(0,0)[c]
{$+i \epsilon_{abc}\vec{p}\cdot \vec{V}^c (\vec{\ell}) $}}
%--------------------- pi-pi -----------------------
\put(75,25){\makebox(0,0)[c]{${\cal H}^{\varphi\chi}_I(\vec{\ell})$}}
\put(67,26){\makebox(0,0)[b]{$\chi$}}
\put(68,24){\makebox(0,0)[tr]{$(p_0,\vec{p})$}}
\put(85,26){\makebox(0,0)[b]{$\varphi_a$}}
\put(83,24){\makebox(0,0)[tl]{$(q_0,\vec{q})$}}
\put(68,25){\circle*{1}}
\put(75,25){\circle{14}}
\put(82,25){\circle*{1}}
\put(75,10){\makebox(0,0)[c]{$\displaystyle
i \vec{q}\cdot\vec{P}^a(\vec{\ell})$}}
%--------------------- chi-pi -----------------------
\put(120,25){\makebox(0,0)[c]{${\cal H}^{\chi\chi}_I(\vec{\ell})$}}
\put(112,26){\makebox(0,0)[b]{$\chi$}}
\put(113,24){\makebox(0,0)[tr]{$(p_0,\vec{p})$}}
\put(129,26){\makebox(0,0)[b]{$\chi$}}
\put(128,24){\makebox(0,0)[tl]{$(q_0,\vec{q})$}}
\put(113,25){\circle*{1}}
\put(120,25){\circle{14}}
\put(127,25){\circle*{1}}
\put(120,10){\makebox(0,0)[c]{$M(\vec{\ell})-m_\chi^2$}}

\end{picture}
\end{center}
\caption{Free propagators and interactions for the pion and the
scalar fields in the presence of a background skyrmion matter. The
energy-momentum conservation $\delta$ functions are not shown. }
\end{figure}
%------------------------ end figure 4 -------------------------

%----------------------------- Figure 5 ------------------------
\begin{figure}
\begin{center}
\setlength{\unitlength}{1mm}
\begin{picture}(150,60)(0,-5)
\linethickness{0.8pt}
%--------------- Sigma1 ---------------------------
\put(30,45){\makebox(0,0)[c]{$\Sigma^{(1)}$}}
\put(20,46){\makebox(0,0)[b]{$\varphi_a$}}
\put(23,44){\makebox(0,0)[tr]{$(p_0,\vec{p})$}}
\put(40,46){\makebox(0,0)[b]{$\varphi_a$}}
\put(38,44){\makebox(0,0)[tl]{$(p_0,\vec{p})$}}
\put(23,45){\circle*{1}}
\put(30,45){\circle{14}}
\put(37,45){\circle*{1}}
%---------------- equal ------------
\put(52,45){\makebox(0,0)[c]{=}}
%--------------- intermediate pions -------------------------
\put(75,45){\makebox(0,0)[c]{${\cal H}^{\varphi\varphi}_I(\vec{0})$}}
\put(65,46){\makebox(0,0)[b]{$\varphi_a$}}
\put(68,44){\makebox(0,0)[tr]{$(p_0,\vec{p})$}}
\put(85,46){\makebox(0,0)[b]{$\varphi_a$}}
\put(83,44){\makebox(0,0)[tl]{$(p_0,\vec{p})$}}
\put(68,45){\circle*{1}}
\put(75,45){\circle{14}}
\put(82,45){\circle*{1}}
%--------------- Sigma2 ---------------------------
\put(30,25){\makebox(0,0)[c]{$\Sigma^{(2)}$}}
\put(20,26){\makebox(0,0)[b]{$\varphi_a$}}
\put(23,24){\makebox(0,0)[tr]{$(p_0,\vec{p})$}}
\put(40,26){\makebox(0,0)[b]{$\varphi_a$}}
\put(38,24){\makebox(0,0)[tl]{$(p_0,\vec{p})$}}
\put(23,25){\circle*{1}}
\put(30,25){\circle{14}}
\put(37,25){\circle*{1}}
%---------------- equal ------------
\put(52,25){\makebox(0,0)[c]{=}}
%--------------- intermediate pions -------------------------
\put(75,25){\makebox(0,0)[c]{${\cal H}^{\varphi\varphi}_I(\vec{\ell})$}}
\put(65,26){\makebox(0,0)[b]{$\varphi_a$}}
\put(68,24){\makebox(0,0)[tr]{$(p_0,\vec{p})$}}
\put(85,26){\makebox(0,0)[b]{$\varphi_b$}}
\put(83,24){\makebox(0,0)[tl]{$(p_0,\vec{p}+\vec{\ell})$}}
\put(68,25){\circle*{1}}
\put(75,25){\circle{14}}
\put(82,25){\circle*{1}}
\multiput(82,25)(4,0){5}{\line(1,0){2.5}}
\put(108,25){\makebox(0,0)[c]{${\cal H}^{\varphi\varphi}_I(-\vec{\ell})$}}
\put(98,26){\makebox(0,0)[b]{$\varphi_b$}}
\put(118,26){\makebox(0,0)[b]{$\varphi_a$}}
\put(116,24){\makebox(0,0)[tl]{$(p_0,\vec{p})$}}
\put(101,25){\circle*{1}}
\put(108,25){\circle{14}}
\put(115,25){\circle*{1}}
%---------------- intermediate chi ------------
\put(52,5){\makebox(0,0)[c]{+}}
\put(75,5){\makebox(0,0)[c]{${\cal H}^{\varphi\chi}_I(\vec{\ell})$}}
\put(65,6){\makebox(0,0)[b]{$\varphi_a$}}
\put(68,4){\makebox(0,0)[tr]{$(p_0,\vec{p})$}}
\put(85,6){\makebox(0,0)[b]{$\chi$}}
\put(83,4){\makebox(0,0)[tl]{$(p_0,\vec{p}+\vec{\ell})$}}
\put(68,5){\circle*{1}}
\put(75,5){\circle{14}}
\put(82,5){\circle*{1}}
\put(82,5){\line(1,0){20}}
\put(108,5){\makebox(0,0)[c]{${\cal H}^{\chi\varphi}_I(-\vec{\ell})$}}
\put(98,6){\makebox(0,0)[b]{$\chi$}}
\put(118,6){\makebox(0,0)[b]{$\varphi_a$}}
\put(116,4){\makebox(0,0)[tl]{$(p_0,\vec{p})$}}
\put(101,5){\circle*{1}}
\put(108,5){\circle{14}}
\put(115,5){\circle*{1}}

\end{picture}
\end{center}
\caption{Diagrams used to evaluate the self-energy of the $\varphi_a$
propagation up to second order in the interaction.
Here, $b$ runs over $1,2,3$ and the intermediate states run
over all $\vec{\ell} \neq 0$.}
\end{figure}
%---------------------------- end figure 5 -------------------------------

We show in Fig.~5 the diagrams used to evaluate the self-energy.
Only the diagrams for $\Sigma_{\varphi_a \varphi_b}$ appear. The
symmetry structure of the skyrmion matter allows a nonvanishing
self-energy only for $a=b$, One easily gets the corresponding
diagrams for $\Sigma_{\varphi_a \chi}$ or $\Sigma_{\chi\chi}$. To
first order, $\Sigma^{(1)}$ is nothing but
${\cal H}_I(\vec{\ell}=\vec{0})$. Since ${\cal H}^{\varphi\chi}(\vec{0})=0$,
no mixing between the fluctuating pions and the fluctuating scalar
occurs. Thus, the pion propagator for $\varphi_a$ can be expressed
as
\begin{equation}
\frac{1}{p_0^2- \vec{p}^2 - \Sigma^{(1)}(p_0,\vec{p})}
= \frac{1}{G^{aa}(\vec{0}) (p_0^2 - \vec{p}^2 ) }.
\label{propagator}\end{equation}
where we have used that the self energy to this order is
$ \Sigma^{(1)}_{\varphi_a\varphi_a}(p_0,\vec{p})=
-p^2 (G^{aa}(\vec{0})-1). $
The superscript ``(1)" means that the quantities are evaluated
to first order.
Comparing these results with the propagator (\ref{prop_eff}), we obtain
$f_t=f_s=f_\pi\sqrt{G^{aa}(\vec{0})}$.
Since $G^{aa}(\vec{0})$ is nothing but the average of
$G^{aa}(\vec{r})$ over the space, our calculation thus far reproduces
the  results of ref.~\cite{LPRV03}.
To the same order, the self-energy of the scalar field
is
$\Sigma^{(1)}_{\chi\chi} = M(\vec{0})-m_\chi^2$.
Since it is constant, this self-energy modifies just the
scalar mass from the free value $m_\chi$ to $\sqrt{M(\vec{0})}$.

Now, let us evaluate the second order diagrams shown in Fig.~5.
Again, the symmetry of the background skyrmion matter does not
allow nonvanishing off diagonal components such as
$\Sigma^{(2)}_{\varphi_a\varphi_b}(a \neq b)$
and $\Sigma^{(2)}_{\chi\varphi_a}$.
We obtain
\begin{eqnarray}
\Sigma^{(2)}_{\varphi_a\varphi_a}(p_0,\vec{p})&=&
\sum_{\vec{\ell}\neq 0} \left\{
\frac{|(p_0^2-\vec{p}\cdot(\vec{p}+\vec{\ell}))
G^{ab}(\vec{\ell})+
\epsilon_{abc}\vec{p}\cdot\vec{V}_c(\vec{\ell})|^2}
{p_0^2-(\vec{p}+\vec{\ell})^2} \right. \nonumber\\
&&\hskip 10em \left.
+ \frac{|\vec{p}\cdot\vec{P}_a(\vec{\ell})|^2}
{p_0^2-(\vec{p}+\vec{\ell})^2-m_\chi^2} \right\},
\label{S2pp} \\
\Sigma^{(2)}_{\varphi_a\chi}(p_0,\vec{p})&=& 0,
\label{S2pc} \\
\Sigma^{(2)}_{\chi\chi}(p_0,\vec{p})&=& \sum_{\vec{\ell}\neq 0}
\left\{ \frac{|\vec{p}\cdot\vec{P}_a(\vec{\ell})|^2}
{p_0^2-(\vec{p}+\vec{\ell})^2} +
\frac{|M(\vec{\ell})|^2}{p_0^2-(\vec{p}+\vec{\ell})^2-m_\chi^2}
\right\}.
\label{S2cc}
\end{eqnarray}
Here and in what follows, we economize on notation, $|O^b|^2 =
\sum_b O^b (O^b)^*$ (also for lower indices), and we sum over
index $c$. Again, in spite of the $\varphi-\chi$ coupling term in
the interaction Lagrangian (\ref{Lint}),
$\Sigma^{(2)}_{\chi\varphi_a}$ vanishes so that the pion
propagator and the scalar propagator can be simply written as
\begin{equation}
\frac{1}{p_0^2 - \vec{p}^2 - \Sigma^{(1+2)}_{\varphi_b\varphi_a}},
\hskip 3em
\frac{1}{p_0^2 - \vec{p}^2 - m_\chi^2 - \Sigma^{(1+2)}_{\chi\chi}},
\end{equation}
respectively.

The $p_0$ and $\vec{p}$ dependence of $\Sigma^{(2)}$ is not so
simple as that of $\Sigma^{(1)}$.
Let's consider the {\em small} energy-momentum region.
If we expand the self-energy $\Sigma_{\varphi_a \varphi_b}$
in powers of $p_0$ and $\vec{p}$, then
the coefficients to second order  in $p_0$ and
$\vec{p}^2$ lead to
\begin{eqnarray}
\left( \frac{f_t}{f_\pi} \right)^2 &=&
1- \frac12  \left. \frac{\partial^2 \Sigma^{(1+2)}_{\varphi_a\varphi_a}}
{\partial p_0^2} \right|_{0,\vec{0}} = G^{aa}(\vec{0}),
\label{zt2}\\
\left( \frac{f_s}{f_\pi} \right)^2 &=& 1 +
 \frac12 \left. \frac{\partial^2 \Sigma^{(1+2)}_{\varphi_a\varphi_a}}
{\partial p_i^2} \right|_{0,\vec{0}}
= G^{aa}(\vec{0})
-\sum_{\vec{\ell}\neq 0}
\left\{
\frac{|V_c^i|^2 + \ell_i^2 |G^{ab}(\vec{\ell})|^2}{\vec{\ell}^2}
+\frac{|P^i_a(\vec{\ell})|^2}{\vec{\ell}^2+m_\chi^2} \right\}.
\label{zs2}
\end{eqnarray}
To obtain this result, we have used the observation that
$\partial^2 \Sigma^{(1+2)}/\partial p_i
\partial p_j$ is diagonal in the limit of
$p_0 \rightarrow 0$ and $\vec{p}\rightarrow 0$. Note that the second
order correction terms contribute only to $f_s/f_\pi$ and since
they are all negative the pion velocity becomes  $v_\pi < 1$.

In a similar way, the scalar mass gets a second order correction given by
\begin{equation}
m_\chi^{*2} = m_\chi^2 + \Sigma^{(1+2)}_{\chi\chi}(0,\vec{0})
= M(\vec{0}) - \sum_{\vec{\ell}\neq 0}
\frac{|M(\vec{\ell})|^2}{\ell^2+m_\chi^2}.
\end{equation}

The results of our calculation are shown in Figs.~6. Both of the
pion decay constants change significantly as a function of density
and vanish -- in the chiral limit -- when chiral symmetry is
restored. However, the second-order contributions to the $f_s$ and
$f_\pi$, which break Lorentz symmetry, turn out to be rather
small, and thus their ratio, the pion velocity, stays $v_\pi \sim
1$. The lowest value found is $\sim 0.9$. Note, however, the
drastic change in its behavior at two different phase transition
densities. At the lower density, where the skyrmion matter is in
the chiral symmetry broken phase, the pion velocity decreases and
has the minimum at $\rho = \rho_p$. If one worked only at low
density in a perturbative scheme, one would conclude that the pion
velocity decreases all the way to zero. However, the presence of
the pseudogap phase transition changes this behavior. In the
pseudogap phase, the pion velocity not only stops decreasing but
also starts increasing with increasing density. In the chiral
symmetry restored phase both $f_t$ and $f_s$ vanish. This result
can be comparable to that of ref.~\cite{HKRS03} found in the heat
bath, where the pion velocity approaches 1 while both the spatial
and temporal pion decay constants vanish at $T=T_c$.

As stressed in Sec. 2, because of the strong parameter dependence,
e.g., on $e$ and $m_\chi$, the precise values at which these
transitions occur may not be meaningful. What is relevant
independently of the precise value of $e$ and $m_\chi$ is the
overall qualitative behavior of the pion velocity as a function of
the density. A qualitative idea can be gained by the rough
estimates: The pseudogap type phase transition occurs at
\begin{equation}
\rho_p \sim 0.004 m_\rho^3 (\sim 1.4 \rho_0
\mbox{  with $m_\rho = 770$ MeV)},
\end{equation}
with the minimum of the pion velocity at $v_\pi\sim 0.9$ and the
chiral phase transition takes place at
\begin{equation}
\rho_c\sim 0.01 m_\chi^2(\sim 2.9\rho_0 \mbox{ with
$m_\chi=720$MeV)},
\end{equation}
where the pion decay constants vanish. For the above estimates, we
have used, for reference, the result of the analysis of the deeply
bound states of pions in heavy nuclei~\cite{Yamazaki02}:
\begin{equation}
(f_\pi^*/f_\pi)^2|_{\rho=0.6 \rho_0} = 0.78 \pm 0.05
\end{equation}
which is consistent with our naive estimates given above.

\begin{figure}
\centerline{\epsfig{file=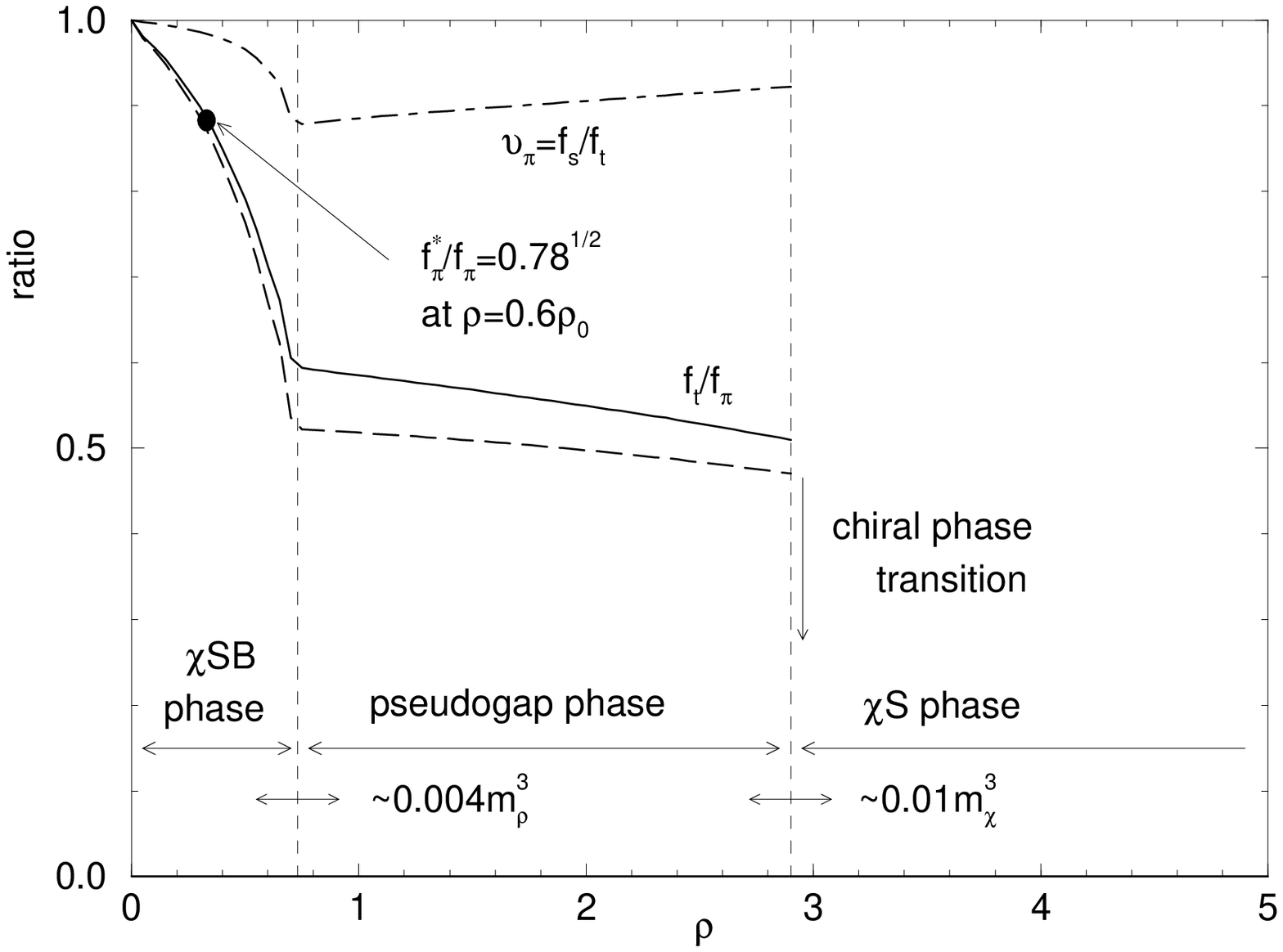,width=8cm,height=11cm,angle=270}} \
\caption{In-medium pion decay constants and their ratio, the pion
velocity.}
\end{figure}

\section{Conclusion}

We have developed a formalism to calculate in-medium properties of
hadrons within a unified approach which describes both matter and
meson fluctuations with the same Lagrangian. In this paper, we
have focused on the pion velocity, which controls through a
dispersion relation, the pion propagation in the medium. For
simplicity, we have chosen massless pions, i.e., the chiral limit
and therefore the propagation is entirely controlled by the
velocity.

A novel aspect that was uncovered here in the skyrmion picture in
the chiral limit is the presence of a psuedogap-like phase. This
phase was not noticed in the previous work~\cite{LPRV03} with
massive pions. As density increases, the system first undergoes a
phase transition to the pseudogap phase where the average value of
$\sigma$ vanishes while the chiral circle preserves a finite
radius. Roughly speaking, chiral symmetry is only partially
restored in this phase. If we increase the density further, the
radius of the chiral circles continuously decreases to zero and
the chiral symmetry is truly restored. The density range where the
skyrmion matter is in such a pseudogap phase depends strongly on
the scalar mass.

Whether or not this pseudogap phase is an artifact of the Skyrme
model at a classical level is not clear. The skyrmion comes out as
a topological object in mapping the $SU(2)$ space defined on the
chiral circle into the space. As a natural consequence, the space
occupied by the skyrmions would have a value for $\langle \sigma
\rangle$ which is smaller than the radius of the chiral circle. It
would be very interesting to check on lattice whether such a
pseudogap phase can be realized in QCD as suggested by Zarembo in
heat bath~\cite{zarembo}.

The behavior of the pion decay constants, $f_t$ and $f_s$,
drastically changes in the various phases of skyrmion matter. The
constants vanish in the chiral symmetry restored phase in the
chiral limit. The pion velocity obtained as the ratio of these two
pion decay constants can deviate at most by 10\% from its vacuum
value 1, since the second order terms of the interaction between
the pion and background matter which break Lorentz symmetry have a
small effect in the decay constants. After achieving its minimum
value at $\rho = \rho_p$, this ratio increases asymptotically to 1
up to the density where the pion velocity is well-defined.

Our model calculation produces a complex phase transition scenario
which seems to differ from the HLS/VM scenario in heat bath~\cite{HKRS03},
despite the fact that the pion velocity ultimately
tends to 1 as in the HLS/VM theory modulo Lorentz symmetry
breaking effect, and also from the linear sigma model in the
$O(4)$ universality class~\cite{SS02a}.

Finally we should point out the short-comings in the calculation
and further works to be done. First of all, ours is not a
realistic calculation in that the model we have used may not be
describing nuclear matter properly. Specifically we do not have
the Fermi liquid structure of normal nuclear matter. Even so, we
do nonetheless expect the qualitative features of our calculation
to be more or less unaffected even after quantum fluctuations are
introduced and the liquid structure is recovered. Next we must
note that our calculation was made to second order in perturbation
theory which may be justified at low densities but we do not have
at present a control of this approximation for larger densities.

Other short-comings are that the kinetic energy contributions are
missing from the calculation~\cite{cohen} and that the subtlety in
the interchange of the chiral limit and infinite volume limit to
address phase transitions needs to be clarified.

The immediate improvements to be made in the calculation are as
follows. Apart from the quantum effects that would bring a liquid
structure to the matter, it would be necessary to assure that the
in-medium ``vacuum" state is stable under mesonic fluctuations.
For instance, we need to examine the eigen-modes and eigenvalues
of eq.(20), focusing on zero-modes and negative modes if any. This
is an interesting issue connected with the possible pion
condensation etc. In this connection, we need to depart from the
chiral limit.

All these short-comings and improvements are under investigation
and will be reported in a later publication.
\section*{Acknowledgements}
Byung-Yoon Park is grateful for the hospitality of CSSM(Center for
the Subatomic Structure of Matter) at the University of Adelaide,
where part of this work has been done. This work was partially
supported by grants MCYT-FIS2004-05616-C02-01 and
GV-GRUPOS03/094~(VV,HJL), KOSEF Grant R01-1999-000-00017-0~(BYP).

\end{document}